\documentclass[a4paper]{jpconf}
\usepackage{graphicx}
\begin{document}
\title{High pressure study on the strong-coupling superconductivity in non-centrosymmetric compound CeIrSi$_3$\footnote{This paper is prepared for ``5th International Conference on Magnetic and Superconducting Materials, MSM07 (25th-30th September 2007, Khiva, Uzbekistan)".}}

\author{Naoyuki Tateiwa$^{1}$, Yoshinori Haga$^{1}$, Tatsuma D. Matsuda$^{1}$, Shugo Ikeda$^{1}$, Etsuji Yamamoto$^{1}$, Yusuke Okuda$^{2}$, Yuichiro Miyauchi$^{2}$, Rikio Settai$^{2}$ and Yoshichika {\=Onuki}$^{1,2}$}

\address{$^1$Advanced Science Research Center, Japan Atomic Energy Agency, Tokai, Ibaraki 319-1195, Japan}
\address{$^2$Graduate School of Science, Osaka University, Toyonaka, Osaka 560-0043, Japan}

\ead{tateiwa.naoyuki@jaea.go.jp}

\begin{abstract}
 We have carried out high pressure experiment on the pressure-induced superconductor CeIrSi$_3$ without inversion center. The electrical resistivity and ac heat capacity were measured in the same run for the same sample. The critical pressure of the antiferromagnetic state was determined to be $P_{\rm c}$ = 2.25 GPa. The heat capacity $C_{\rm ac}$ shows both antiferromagnetic and superconducting transitions at pressures close to $P_{\rm c}$. The co-existence of the antiferromagnetism and superconductivity is discussed. The superconducting region is extended up to about 3.5 GPa. The superconducting transition temperature $T_{\rm sc}$ shows a maximum value of 1.6 K around $2.5-2.7$ GPa. At 2.58 GPa, a large heat capacity anomaly was observed at $T_{\rm sc}$ = 1.59 K. The jump of the heat capacity in the form of ${\Delta}{C_{\rm ac}}/C_{\rm ac}(T_{\rm sc})$ is 5.7 $\pm$ 0.1. This is the largest value observed among all superconductors studied previously, suggesting the strong-coupling superconductivity in CeIrSi$_3$. The large magnitude and anisotropy of the upper critical field $B_{\rm c2}$ at 2.65 GPa is discussed from view points of the strong-coupling superconductivity and the reduced paramagnetic effect in the non-centrosymmetric superconductor. Above $P_{\rm c}$, the electrical resistivity shows the anomalous $T$-linear dependence in the wide temperature region from $T_{\rm sc}$ to 30 K, which is different from the Fermi liquid theory. Meanwhile, the heat capacity $C_{\rm ac}/T$ shows a simple temperature dependence in the normal state above $T_{\rm sc}$. These features do not seem to be explained simply by the spin fluctuation theory. The electronic specific heat coefficient at $T_{\rm sc}$ is approximately unchanged as a function of pressure, even at $P_{\rm c}$. The superconductivity in CeIrSi$_3$ may be different from those appeared around the magnetic instability.

\end{abstract}

\section{Introduction}
  Recently, the discovery of non centrosymmetric superconductors such as CePt$_3$Si, UIr, CeIrSi$_3$, CeRhSi$_3$, and CeCoGe$_3$ has attracted considerable interest~\cite{bauer,akazawa,sugitani,kimura,settai}. In a centrosymmetric compound, conduction bands are degenerate with respect to the ``spin'' degree of freedom. But, in a non centrosymmetric compound, degenerate bands are split due to the Rashba-type spin-orbit interaction, which has strong influence the superconducting properties, particularly the pairing symmetry of the Cooper pairs~\cite{edelstein,gorkov,frigeri,samokhin,sergienko,mineev,fujimoto1,fujimoto2,fujimoto3}. It was revealed theoretically that a mixed-type pair wave function with spin-singlet and spin-triplet components is realized in a non-centrosymmetric superconductor. In CePt$_3$Si, the two-component order parameter of the superconducting pair wave function was suggested in the recent NMR experiment~\cite{yogi}. Many theoretical and experimental studies have been extensively conducted in order to clarify this novel type of unconventional superconductivity. 
 
  In this paper, we report our high pressure study on the pressure-induced superconductivity CeIrSi$_3$. The Ce-based 1-1-3 system, CeTX$_3$ (T: Transition metal, X: Si, Ge) has been systematically investigated\cite{muro}. The ground state of the system varies from the magnetic Kondo lattice to the non-magnetic state through the heavy-fermion state by the replacement of the transition metal or Si (Ge) element. The 1-1-3 system crystalizes in the tetragonal BaNiSn$_3$-type crystal structure ($I$4$mm$).  The Ce atoms occupy the four corners and the body center of the tetragonal structure in a manner similar to the well-known tetragonal ThCr$_2$Si$_2$-type structure which many heavy fermion superconductors such as CeCu$_2$Si$_2$, CeCu$_2$Ge$_2$, CePd$_2$Si$_2$, CeRh$_2$Si$_2$ and URu$_2$Si$_2$ belong to\cite{steglich,jaccard, mathur,movshovich,palstra1}. Both BaNiSn$_3$-type and ThCr$_2$Si$_2$-type structures are a derivative of the BaAl$_4$-type structure. There is no inversion center in CeIrSi$_3$ due to the asymmetric arrangements of the Ir and Si atoms. The point group of CeIrSi$_3$ is $C_{4v}$ that lacks a mirror plane and a two-fold axis normal to the $c$-axis. A potential gradient ${\nabla}V(\mathbf{r})$ appears along the $c$ axis. Here, $V(\mathbf{r})$ is the periodic potential of the crystal lattice. The Fermi surface properties of the 1-1-3 have been investigated by the de Hass-van Alphen (dHvA) experiments on CeRhSi$_3$, CeCoGe$_3$, LaCoGe$_3$, and LaIrSi$_3$\cite{kimura2,thamizhavel,okuda}. For LaIrSi$_3$, the Fermi surface is found to split into two Fermi surfaces due to the spin-orbit interaction arising from the non-centrosymmetric crystal structure. The separation energy is in the range of 95-1100 K which is two orders of magnitude larger than the superconducting energy gap\cite{okuda}. 

    At ambient pressure, the electrical resistivity of CeIrSi$_3$ shows a broad resistivity shoulder around 100 K, which is a characteristic feature of the CeTX$_3$ system related to the combined effect of the Kondo effect and the crystalline electric field (CEF) effect\cite{muro}. The Kondo temperature $T_{\rm K}$ is estimated to be about 100 K. CeIrSi$_3$ shows an antiferromagnetic transition at a N\'{e}el temperature $T_{\rm N}$ = 5.0 K. The magnetic entropy $S_{\rm mag}$ is 0.2$R\,$ln$\,$2 at $T_{\rm N}$. This small value suggests the itinerant character of the 4$f$ electron in CeIrSi$_3$ due to the Kondo effect. The appearance of an internal field in the antiferromagnetic state is recently confirmed by the muon spin rotation ($\mu$SR) experiment~\cite{higemoto}. The size of the magnetic moment in the ordered state is estimated roughly as ${\mu}_{\rm ord}$ = $0.3-0.5\,{\mu}_{\rm B}$/Ce. Under high pressure, the N\'{e}el temperature decreases monotonically with increasing pressure and disappears at around 2 GPa. Superconductivity appears in a wide pressure region from 1.7 GPa to about 3.5 GPa, with a relatively large superconducting transition temperature $T_{\rm sc}$ = 1.6 K around 2.5 GPa~\cite{sugitani,okuda}. The large value of the slope of the upper critical field $-{\rm d}B_{\rm c2}/{\rm d}T_{\rm sc}$ at $T_{\rm sc}$ suggests the superconductivity of heavy quasiparticles. A characteristic feature of the superconducting state in CeIrSi$_3$ is the large magnitude and anisotropy of the upper critical field $B_{\rm c2}(T)$. The value of $B_{\rm c2}(T)$ for $B$ $\parallel$ [001] at 2.65 GPa is extremely high, roughly estimated as 30 T at 0 K. It is noted that a large value of $B_{\rm c2}(T)$ is also reported in CeRhSi$_3$\cite{kimura3}. 
   
   We performed the ac heat capacity and electrical resistivity measurements on CeIrSi$_3$ in order to study the superconductivity and antiferromagnetism. We show that the strong-coupling superconductivity is realized in this compound and various superconducting properties are discussed on this point of view.  The co-existence of the antiferromagnetism and superconductivity is considered. We will discuss the pressure change of the electronic state around the magnetic critical pressure $P_{\rm c}$. The non-Fermi liquid behavior was observed above $P_{\rm c}$, which will be considered from various theoretical points of views. The large magnitude and anisotropy of the upper critical field $B_{\rm c2}$(0) is analyzed by the strong-coupling model and the theoretical prediction for the non-centrosymmetric superconductor.

 \section{Experiment}
 The single crystal of CeIrSi$_3$ was grown by the Czochralski method in a tetra-arc furnace. The details of the sample preparation are given in the recent paper\cite{okuda}. The residual resistivity ratio RRR (= ${\rho}_{\rm RT}/{\rho}_{\rm 0}$) is 120, where ${\rho}_{\rm RT}$ and ${\rho}_{\rm 0}$ are the resistivity at room temperature and the residual resistivity, respectively, indicating the high quality of the sample. The heat capacity under high pressures was measured by the ac calorimetry method~\cite{wilhelm,tateiwa}.  The sample was heated up using a heater, whose power is modulated at a frequency ${\omega}$. The amplitude of the temperature oscillation $T_{\rm ac}$ is written as a function of heat capacity $C_{ac}$ of the sample $ T_{\rm ac} = P_{0}/({\kappa}+i{\omega}C_{ac})$. Here, $P_{0}$ is an average of the power. ${\kappa}$ is the thermal conductivity between the sample and the environment. $T_{\rm ac}$ was measured with a AuFe/Au thermocouple (Au + 0.07 at\% Fe). The contribution from the thermocouple and Au wires to the heat capacity is very small ($\sim$ 0.1\%).  The resistivity measurement was also carried out for the same sample by the standard four-terminal method.  For the resistivity measurement, two additional Au wires were attached to the edges of the sample so as to pass the electrical current. The low-temperature measurement was carried out using a $^{3}$He refrigerator from 0.3 K to 10 K. We used a hybrid piston cylinder-type cell. Daphne oil (7373) was used as a pressure transmitting medium~\cite{uwatoko}.

    \begin{figure}[t]
\begin{center}
\includegraphics[width=15cm]{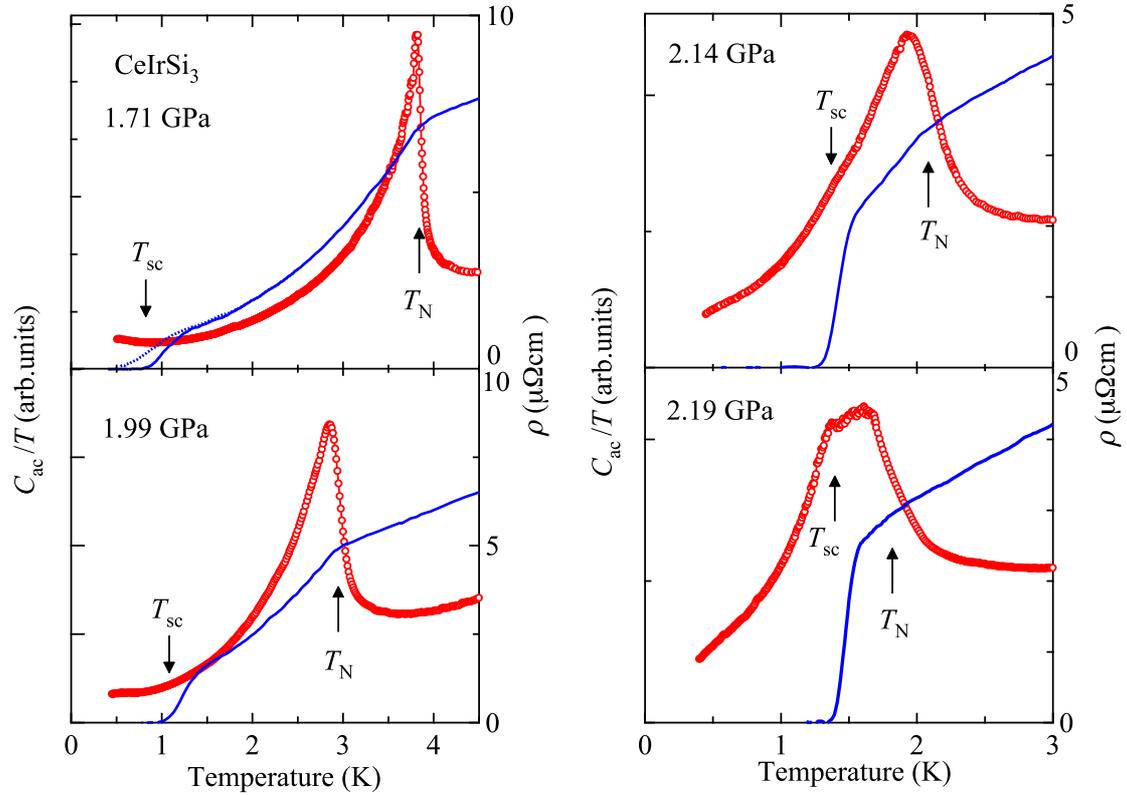}
 \end{center}
\caption{\label{fig:epsart} Temperature dependences of the ac heat capacity $C_{ac}$ (left side) and electrical resistivity $\rho$ (right side) at 1.71, 1.99, 2.14 and 2.19 GPa in CeIrSi$_3$.}
\end{figure}

\section{Results and Discussions}
\subsection{Heat capacity and electrical resistivity at low temperatures}
   Figure 1 shows the temperature dependences of the heat capacity $C_{\rm ac}$ and electrical resistivity $\rho$ at 1.71, 1.99, 2.14 and 2.19 GPa, below the critical pressure $P_{\rm c}$ = 2.25 GPa.  At 1.71 GPa, $C_{\rm ac}$ shows a clear anomaly and $\rho$ does a kink at the N\'{e}el temperature $T_{\rm N}$ = 3.88 K. At low temperatures, the behavior of $\rho$ depends on the applied electrical current $j$. The data for $j$ = 1.0 and 0.3 A/cm$^2$ are shown by the dotted and solid lines, respectively. The resistivity reveals a superconducting transition at $T_{\rm sc}$ = 0.85 K and 0.55 for $j$ = 0.5 A/cm$^2$ and $j$ = 1.0 A/cm$^2$, respectively. However, $C_{\rm ac}$ does not show an anomaly at $T_{\rm sc}$.  At 1.99 GPa, $C_{\rm ac}$ and $\rho$ show a clear transition at $T_{\rm N}$ = 2.98 K, and only resistivity reveals a superconducting transition at $T_{\rm sc}$ = 1.02 K. No evidence of the bulk superconductivity is obtained at 1.31 (data not shown), 1.71, and 1.99 GPa. At 2.14 GPa, both $\rho$ and $C_{\rm ac}$ show the clear antiferromagnetic transition. $\rho$ shows the superconducting transition at $T_{\rm sc}$ = 1.32 K, and $C_{\rm ac}$ shows a weak hump around $T_{\rm sc}$. At 2.19 GPa, $C_{\rm ac}$ shows a broad anomaly with two peak structures, which correspond to the antiferromagnetic and superconducting transitions, respectively. The N\'{e}el temperature is determined as $T_{\rm N}$ = 1.88 K from the entropy balance. The peak of the heat capacity at the lower temperature side is close to the superconducting transition at $T_{\rm sc}$ = 1.40 K, where $\rho$ becomes zero. 

 \begin{figure}[t]
\begin{center}
\includegraphics[width=15cm]{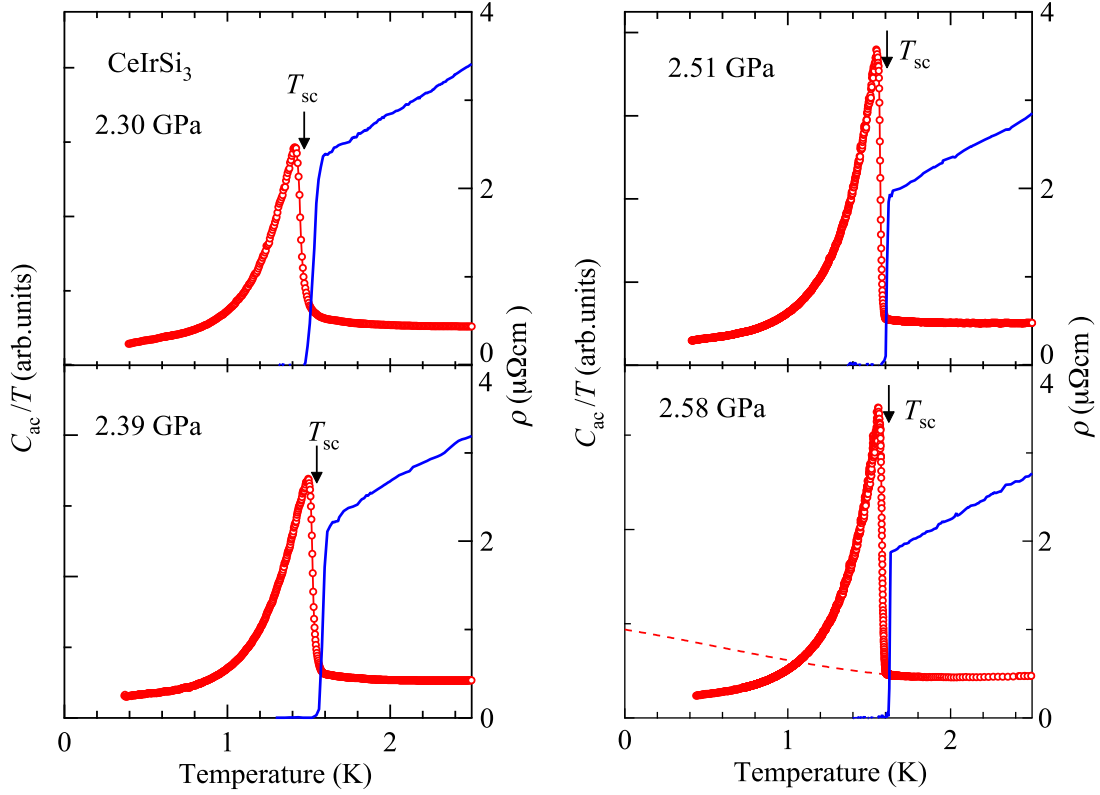}
 \end{center}
\caption{\label{fig:epsart} Temperature dependences of the ac heat capacity $C_{ac}$ (left side) and electrical resistivity $\rho$ (right side) at 2.30, 2.39 and 2.58 GPa in CeIrSi$_3$.}
\end{figure}

  At pressures higher than $P_{\rm c}$ = 2.25 GPa, only the superconducting transition is observed in both $C_{\rm ac}$ and $\rho$, as shown in Figure 2. At 2.58 GPa, the values of $T_{\rm sc}$ are 1.62 and 1.59 K which are obtained from the resistivity and ac heat capacity measurements, respectively. The jump of the heat capacity in the form of ${\Delta}{C_{\rm ac}}/C_{\rm ac}(T_{\rm sc})$ is 3.4 $\pm$ 0.3 at 2.30 GPa and 5.7 $\pm$ 0.1 at 2.58 GPa. Here, $\it{\Delta{C_{\rm ac}}}$ is the jump of the heat capacity at $T_{\rm sc}$ and $C_{\rm ac}(T_{\rm sc})$ is the value of $C_{\rm ac}$ just above $T_{\rm sc}$, namely, corresponding to ${\gamma}T_{\rm sc}$, where ${\gamma}$ is the electronic specific heat coefficient. The values of $\it{\Delta{C_{\rm ac}}/C_{\rm ac}(T_{\rm sc})}$ are extremely larger than the BCS value of 1.43. Especially, the value of 5.7 $\pm$ 0.1 at 2.58 GPa is the largest value among all superconductors previously reported. Here, we considered the entropy balance in the superconducting state of 2.58 GPa, as shown by the dotted line. The value of $C_{\rm ac}/T$ is enhanced with decreasing temperature. The value of $C_{\rm ac}/T$ at 0 K is roungly twice larger than that at $T_{\rm sc}$ = 1.59 K. If the $C_{\rm ac}/T$ value at 0 K is used as the $\gamma$ value,  ${\Delta}{C_{\rm ac}}/({\gamma}T_{\rm sc})$ is about 2.8 $\pm$ 0.3. This is still larger than the BCS value.

     \begin{figure}[t]
\begin{center}
\includegraphics[width=9cm]{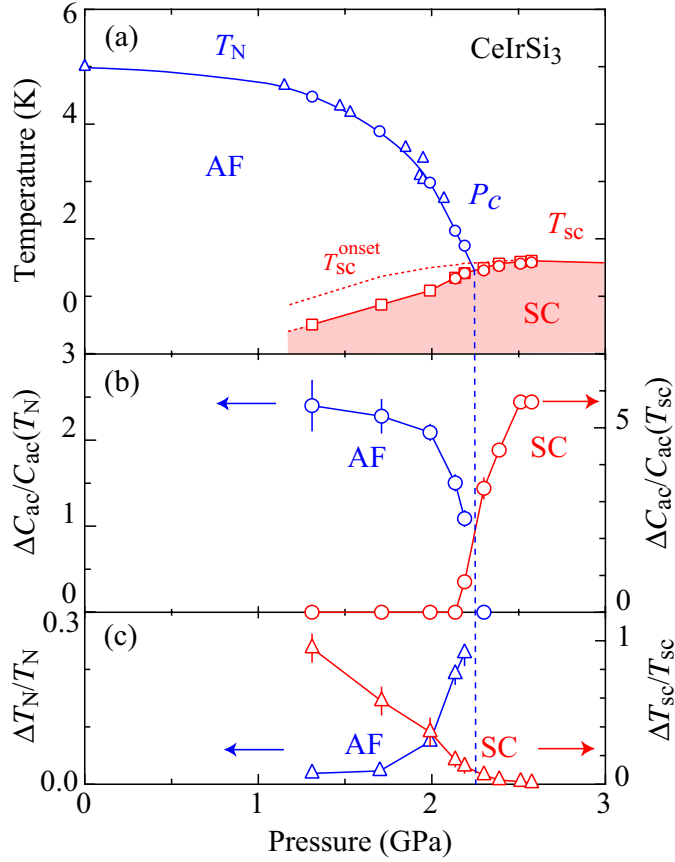}
 \end{center}
\caption{\label{fig:epsart} (a) Pressure phase diagram in CeIrSi$_3$. $T_{\rm N}$, which was determined by the previous resistivity measurements, is plotted by triangles~\cite{sugitani,okuda}. $T_{sc}$ and $T_{\rm N}$ values obtained by the present resistivity and ac heat capacity measurements are shown by squares and circles, respectively. The dotted line indicates the onset temperature of the superconducting transition in the resistivity. (b) Pressure dependences of the jump of the heat capacity anomaly at  $T_{\rm N}$ ${\Delta}{C_{\rm ac}}/C_{\rm ac}(T_{\rm N})$ (left side) and at $T_{\rm sc}$ ${\Delta}{C_{\rm ac}}/C_{\rm ac}(T_{\rm sc})$ (right side). (c) Pressure dependences of the width of the antiferromagnetic transition in the heat capacity ${\it{\Delta}}T_{\rm N}$/$T_{\rm N}$(left side) and the superconducting transition in the resistivity ${\it{\Delta}}T_{\rm sc}$/$T_{\rm sc}$ (right side).}
\end{figure} 

 The absolute value of the heat capacity is not obtained, but the relative change of the heat capacity can be estimated in the ac heat capacity measurement~\cite{wilhelm,tateiwa}. The value of $C_{\rm ac}/T$ just above $T_{\rm sc}$ is determined as 100 $\pm$ 20 mJ/K${^2}{\cdot}$mol at 2.58 GPa by comparison with the value of $C_{\rm ac}$ at ambient pressure. This $\gamma$ value indicates that the moderate heavy-fermion superconductivity is realized in CeIrSi$_3$. This value is approximately the same as $\gamma$ = 120 or 105 mJ/K${^2}{\cdot}$mol at ambient pressure~\cite{muro,okuda}.

 \subsection{Superconductivity and antiferromagnetism}       
   Figure 3(a) shows the pressure phase diagram determined from the present experiment in combination with the previous experimental results~\cite{sugitani,okuda}. $T_{\rm N}$, determined by the previous resistivity measurement, are plotted by triangles~\cite{sugitani,okuda}. $T_{\rm sc}$ and $T_{\rm N}$ values obtained by the present resistivity and ac heat capacity measurements are plotted by squares and circles, respectively. The critical pressure for the antiferromagnetic state $P_{\rm c}$ was determined as $P_{\rm c}$ = 2.25 GPa. In the previous study, the superconductivity was observed above 1.8 GPa, while it is observed in the present resistivity measurement at 1.31 GPa. The reason of this discrepancy is not clear at present.
  \begin{figure}[t]
\begin{center}
\includegraphics[width=9cm]{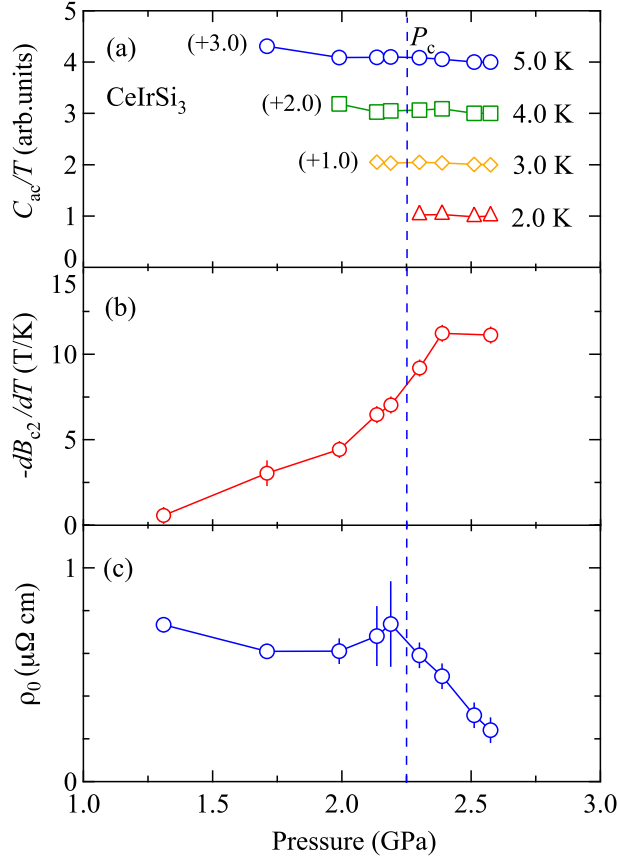}
 \end{center}
\caption{\label{fig:epsart} Pressure dependences of (a) $C_{\rm ac}/T$ at several temperatures, and (b) the slope of the upper critical field -d$H_{\rm c2}/dT$ at $T_{\rm sc}$, and (c) the residual resistivity $\rho_0$ in CeIrSi$_3$.}
\end{figure}

  Figure 3(b) shows the pressure dependence of the jump of the heat capacity anomaly at the antiferromagnetic and superconducting transition temperatures, ${\Delta}{C_{\rm ac}}/C_{\rm ac}(T_{\rm N})$ and  ${\Delta}{C_{\rm ac}}/C_{\rm ac}(T_{\rm sc})$. Figure 3 (c) shows the pressure dependence of the width of the antiferromagnetic transition in the heat capacity ${\it{\Delta}}T_{\rm N}$/$T_{\rm N}$ and the superconducting transition in the resistivity ${\it{\Delta}}T_{\rm sc}$/$T_{\rm sc}$. With increasing pressure above 2 GPa, ${\Delta}{C_{\rm ac}}$/$C_{\rm ac}(T_{\rm N})$ decreases strongly and the antiferromagnetic transition width ${\it{\Delta}}T_{\rm N}$/$T_{\rm N}$ becomes larger. Meanwhile, the jump of the heat capacity anomaly at $T_{\rm sc}$ ${\Delta}{C_{\rm ac}}/C_{\rm ac}(T_{\rm sc})$ starts to increase above 2 GPa and the transition width ${\it{\Delta}}T_{\rm sc}$/$T_{\rm sc}$ becomes small as a function of pressure and close to zero above $P_{\rm c}$ = 2.25 GPa. 
      
  The relation between the antiferromagnetism and superconductivity is the most interesting issue to be discussed. From the present experimental results shown in Figure 3, the superconductivity and antiferrromagnetism in CeIrSi$_3$ seem to be competing with each other. We suggest that the superconductivity and antiferromagnetism do not coexist essentially and the superconductivity exists inhomogeneously below $P_{\rm c}$. It is noted that the pressure dependence of ${\it{\Delta}}T_{\rm sc}$/$T_{\rm sc}$ as well as the gradual increase of ${\Delta}{C_{\rm ac}}/C_{\rm ac}(T_{\rm sc})$ around $P_{\rm c}$ can be interpreted as the increment of the superconducting volume fraction. At present stage, we can not deny the co-existence of both phases completely since the heat capacity $C_{\rm ac}$ shows both antiferromagnetic and superconducting transitions at pressures close to $P_{\rm c}$. However, we suppose that the homogenous co-existence of both phases is not likely.  The antiferromagnetic transition width ${\it{\Delta}}T_{\rm N}$/$T_{\rm N}$ becomes larger in the pressure region close to $P_{\rm c}$. The antiferromagnetic phase may be also spatially inhomogeneous. The absence of the clear heat capacity anomaly at $T_{\rm sc}$ in the antiferromagnetic ordered state might  be explained assuming the homogenous gapless superconductivity which was proposed for CeCu$_2$Si$_2$ and CeRhIn$_5$~\cite{fuseya,ishida,mito}. However, the disappearance of the superconductivity at higher electrical current  $j$ at 1.31 and 1.71 GPa can not be explained by the theory. Rather, it seems to be reasonable to consider an imhomogenous superconducting phase at these pressures. For further investigations on the co-existence of antiferromagnetism and superconductivity, microscopic experiments such as NMR are needed.  
   
    The competent relation between the antiferromagnetism and superconductivity, clarified by the present study in CeIrSi$_3$, seems to be a common feature in the Ce-based superconducting materials such as CeCu$_2$Si$_2$~\cite{thalmeier,flouquet}. However, this feature does not apply to the case of the proto-type non-centrosymmetric superconductor CePt$_3$Si, where the superconductivity appears at $T_{\rm sc}$ = 0.75 K in the antiferromagnetic ordered state below $T_{\rm N}$ = 2.2 K~\cite{bauer}. Our previous high pressure study clarified that the value of ${\Delta}{C_{\rm ac}}/C_{\rm ac}(T_{\rm sc})$ starts to decrease and the superconducting transition width ${\it{\Delta}}T_{\rm sc}$/$T_{\rm sc}$ starts to increase above the critical pressure $P_{\rm c}$ $\sim$ 0.6 GPa~\cite{tateiwa,nicklas}. This suggests the cooperative relation between the two states in CePt$_3$Si, which is very rare case in the Ce-based superconductors. 
        
    The large value of the superconducting heat capacity anomaly ${\Delta}{C_{\rm ac}}/C_{\rm ac}(T_{\rm sc})$ above 2.5 GPa suggests that the strong-coupling superconductivity is realized in CeIrSi$_3$.     
The large jump of  the heat capacity at $T_{\rm sc}$ was also observed in heavy-fermion superconductors CeCoIn$_5$ and UBe$_{13}$ where the value of ${\it{\Delta}}C/({\gamma}T_{\rm sc})$ are 4.5 and 2.7, respectively~\cite{petrovic,ikeda,thomas}. The value at 2.58 GPa in CeIrSi$_3$ is the largest among previously reported superconductors. The strong-coupling effect on superconducting properties have been studied from the theoretical points of view for the $s$-wave superconductor by the electron-phonon interaction or $d$-wave one by the antiferromagnetic spin fluctuations~\cite{scalapino,carbotte,bulaevskii}. The increment of ${\Delta}{C_{\rm ac}}/C_{\rm ac}(T_{\rm sc})$ suggests that the superconducting coupling parameter increases with increasing pressure.

 \subsection{Pressure change of the electronic state around the critical pressure $P_{\rm c}$}     
  Figure 4 (a) shows the pressure dependences of $C_{\rm ac}/T$ at low temperatures. The data in the paramagnetic state are shown, which are normalized by the value at 2.58 GPa and are shifted upwards by one, two, and three scales for 3.0, 4.0, and 5.0 K, respectively, as compared to the data for 2.0 K. There is no distinct change in the pressure dependence of $C_{\rm ac}/T$. This indicates that the electronic specific heat coefficient $\gamma$ is almost pressure independent, even at  the antiferromagnetic critical pressure $P_{\rm c}$ = 2.25 GPa. It is interesting to note that no enhancement is observed in the pressure dependences of cyclotron effective masses in the dHvA experiment on the isostructural pressure-induced superconductor CeRhSi$_3$~\cite{terashima}. Figure 4 (b) the slope of the upper critical field $B_{\rm c2}$ at $T_{\rm sc}$ which is determined from the resistivity measurement in magnetic field along the [110] direction. It becomes large : $-{\rm d}B_{\rm c2}/{\rm d}T$ = 11.2 T/K at 2.58 GPa. In the weak coupling limit,  $-{\rm d}B_{\rm c2}/{\rm d}T$ at $T_{\rm sc}$ is proportional to the square of the effective mass of the conduction electrons, $m^{*2}$~\cite{orlando}. The large value of  $-{\rm d}B_{\rm c2}/{\rm d}T$ = 11.2 T/K at 2.58 GPa is not explained by the existence of conduction electrons with the large effective mass because the $\gamma$ ($\propto$ $m^{*}$) value is approximately unchanged as a function of pressure.  Therefore, the large value of $-{\rm d}B_{\rm c2}/{\rm d}T$ at $T_{\rm sc}$ may be related to the enhancement of the superconducting coupling parameter~\cite{scalapino,carbotte,bulaevskii}. Figure 5 (c) shows the pressure dependence of the residual resistivity $\rho_{0}$, which is almost constant below $P_{\rm c}$ = 2.25 GPa and starts to decrease considerably above $P_{\rm c}$. The value of $\rho_{0}$ at 2.58 GPa (0.24 ${\mu}{\Omega}{\cdot}$cm) is about 25 ${\%}$ of that at ambient pressure (0.96 ${\mu}{\Omega}{\cdot}$cm). 
   
     In pressure-induced superconductors such as CeIn$_3$ or CePd$_2$Si$_2$, the superconductivity appears around the magnetic critical pressure $P_{\rm c}$ where the antiferromagnetic transition temperature $T_{\rm N}$ becomes 0 K~\cite{jaccard,mathur}. The coefficient of $T^2$ term of the resistivity $A$ or the residual resistivity $\rho_0$ show an anomalous enhancement around $P_{\rm c}$. The enhancement is understood as the effect of the critical antiferromagnetic fluctuations around QCP~\cite{moriya,miyake0}. However, in CeIrSi$_3$, there is no anomalous behaviors in the pressure dependences of the $\gamma$ value and $\rho_0$ around the critical pressure $P_{\rm c}$. One possibility is that $P_{\rm c}$ is not second order quantum critical point. It is interesting to note that no anomalous enhancement was observed in the pressure dependence of the $\gamma$ value around the antiferromagnetic critical pressure $P_{\rm c}\sim$ 0.6 GPa of CePt$_3$Si.  Superconductivity in the non-centrosymmetric CeIrSi$_3$ and CePt$_3$Si may be different from superconductivity associated with the magnetic instability around the magnetic critical region.

    \begin{figure}[t]
\begin{center}
\includegraphics[width=16cm]{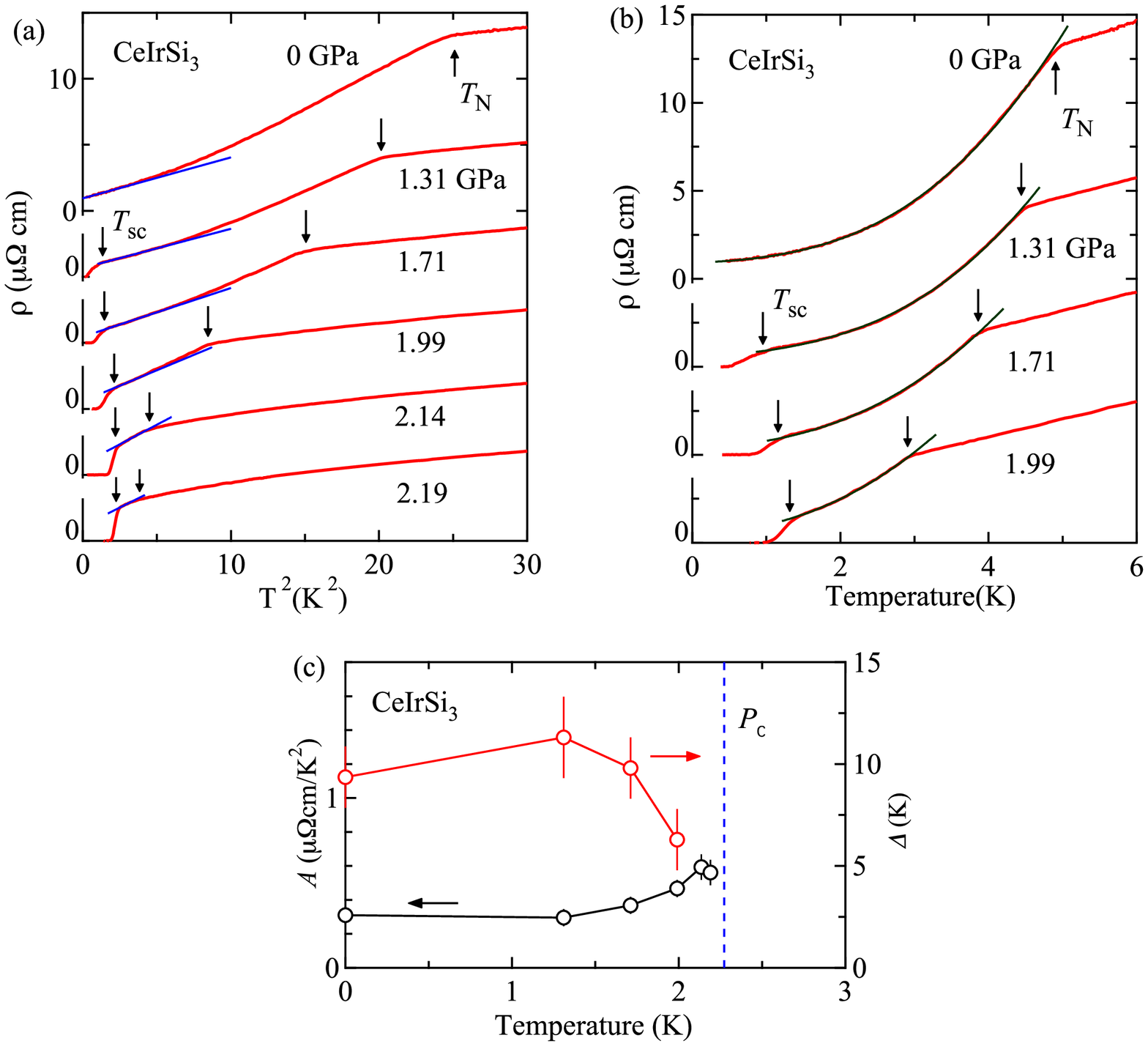}
 \end{center}
\caption{\label{fig:epsart} Electrical resistivity as functions of (a) $T^2$ and (b) $T$ at several pressures below the magnetic critical pressure $P_{\rm c}$ in CeIrSi$_3$. The arrows at the higher and lower temperatures indicate $T_{\rm N}$ and $T_{\rm sc}$, respectively. Blue lines in (a) indicate a Fermi-liquid relation, and green line in (b) corresponds to the fitting curve described in the text. (c) Pressure dependences of the coefficient of $T^2$ term of the resistivity $A$ and the gap of the antiferromagnetic spin wave dispersion $\Delta$. }
\end{figure}  
   
\subsection{Physical properties of the normal state} 

   Figure 5 shows the electrical resistivity as functions of (a) $T^2$ and (b) $T$ below the magnetic critical pressure $P_{\rm c}$. The arrows at the higher and lower temperatures indicate $T_{\rm N}$ and the onset of the superconducting transition temperature $T_{\rm sc}$, respectively.  At low temperatures, the resistivity follows the Fermi-liquid relation ($\rho$ = ${\rho}_0$+$A$$T^2$). Here, the first term ${\rho}_0$ corresponds to the residual resistivity and the second term corresponds to the Fermi-liquid contribution of heavy quasiparticles. The $A$ value is obtained by the fit of the data shown as blue lines in Fig. 5(a).  The resistivity between $T_{\rm sc}$ and $T_{\rm N}$ is analyzed by the antiferromagnetic magnon model which is described as $\rho$ = ${\rho}_0$ + $AT^2$+$BT$(1+2$T/{\Delta}$)exp($-{\Delta}/T$). The third term described the contribution from the scattering by the antiferromagnetic magnon with an energy gap ${\Delta}$.  It is noted that this expression was used in the same context for URu$_2$Si$_2$ and CePd$_2$Si$_2$~\cite{palstra2,raymond}.  A fit of the data is shown as blue line in Fig. 5 (b).  The pressure dependences of $A$ and ${\Delta}$ are shown in Figure 5(c). The value of ${\Delta}$ decreases with increasing pressure roughly above 1.5 GPa but the ratio of ${\Delta}$ and $k_{\rm B}T_{\rm N}$ is almost pressure-independent, indicating that the anisotropy of the antiferromagnetic state does not change significantly under high pressure. At 2.14 and 2.19 GPa, we could not estimate the contribution from the antiferromagnetic magnon to the resistivity uniquely since $T_{\rm N}$ is close to $T_{\rm sc}$. Therefore, the data were analyzed assuming that the resistivity shows the $T^2$-dependence from $T_{\rm sc}$ to $T_{\rm N}$. The obtained $A$ values at two pressures contain contributions not only from the electron-electron scattering but also from the scattering by the antiferromagnetic magnon. It is supposed that the weak enhancement of $A$ at high pressures  is due to the inclusion of the electron-magnon interaction. 
   
      \begin{figure}[t]
\begin{center}
\includegraphics[width=16cm]{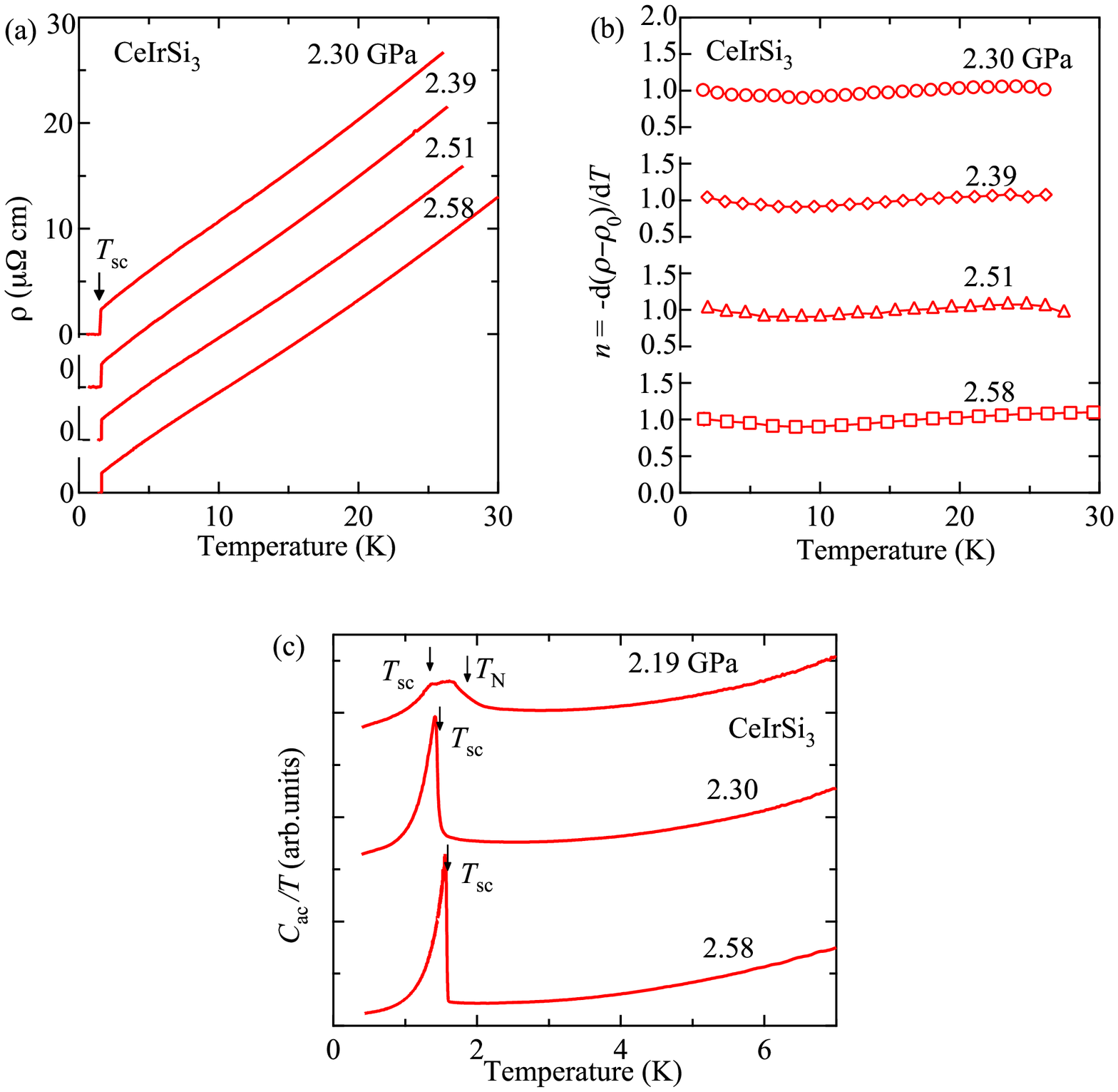}
 \end{center}
\caption{\label{fig:epsart} (a) Electrical resistivity as a function of $T^2$ for the pressure region above the critical pressure $P_{\rm c}$ in CeIrSi$_3$. (b) Temperature dependence of the resistivity exponent $n$ = $-{\rm d}$ln(${\rho}-{\rho}_{0}$)/${\rm d}$ln$T$. (c)Temperature dependence of the heat capacity $C_{\rm ac}/T$ at 2.30 and 2.58 GPa. Experimental data at 2.19 and 2.30 GPa are sifted upwards}
\end{figure} 

   Figure 6 (a) shows the temperature dependence of the electrical resistivity $\rho$ in the pressure region above $P_{\rm c}$. The resistivity shows almost $T$-linear dependence up to 20 K, which is different from the conventional Fermi liquid behavior ($\rho$ = ${\rho}_0$+$A$$T^2$). Figure 6 (b) shows the temperature dependence of  the resistivity exponent $n$ = $-{\rm d}$ln(${\rho}-{\rho}_{0}$)/${\rm d}$ln$T$. The temperature dependences of $\rho$ are almost same at the pressures investigated. The value of $n$ is 1.0 at 20 K and decreases weakly with decreasing temperature, shows a broad minimum  of 0.9 around 7 K and then saturate to 1.0 at low temperatures. 

   In a number of the heavy-fermion compounds on the border to magnetism, the electrical resistivity shows unusual behavior ($\rho{\,}\propto{\,}T^x$ with $x{\,}<{\,} 2$) which is different from the Fermi- liquid theory. This is evidence for an anomalous quasiparticle scattering mechanism. The spin-fluctuation theory predicts a dependence of the resistivity around the magnetic instability as $\rho{\,}\sim{\,}T^{d/z}$, where $d$ is the dimensionality of the spin flucuation spectrum and $z$ is the dynamical exponent which is normally taken to be 2 for the case of an antiferromagnet~\cite{moriya,millis,lonzarich}. Thus, one would expect to observe the exponent $n$ = 1 for the two-dimensional antiferromagnetic system.  In the case of CeIrSi$_3$, however, the anisotropy of the magnetization in the antiferromagnetic state is not large at ambient pressure and ${M_{[100]}}/{M_{[001]}}$ is at most 2 at 1.8 K where ${M_{[100]}}/{M_{[001]}}$ indicates the ratio of the magnetizations for $B$ $\parallel$ [100] and [001] at low magnetic field. Also, there is no strong low-dimensional character in the Fermi surface topology of the 1-1-3 system~\cite{kimura2,thamizhavel,okuda}.  Furthermore, the three-dimensional character of the antiferromagnetic spin fluctuations was recently suggested in the NMR experiment at 2.6 GPa~\cite{kitaoka}, which is contrary to the theoretical expectation. The theory also predicts the anomalous behavior of the heat capacity around the magnetic quantum critical point, $C/T$  $\propto$ -$lnT$ and $T^{1/2}$ for two- and three-dimensional antiferromagnets, respectively~\cite{moriya}. However, there is no anomalous behavior in the temperature dependences of $C_{\rm ac}/T$ above $T_{\rm sc}$ at 2.19, 2.30 and 2.58 GPa as shown in Figure 6 (c). $C_{\rm ac}/T$ show a monotonic and weak temperature dependence in the normal state. It seems that the anomalous physical properties of the normal state is not simply explained by the spin fluctuation theory.  

    On the different point of view, Hlubina and Rice showed that the resistivity of a clean metal close to an antiferromagnetic quantum critical point is dominated by quasiparticles from regions of the Fermi surface far away from the ``hot line'' (points at the Fermi surface connected by the ordering wave vector $Q$) and accordingly, the resistivity shows the $T^2$ dependence~\cite{hlubina}.  Rosch studied the effect of the weak isotropic impurity scattering on the scenario by Hlubina and Rice and showed that the exponent $n$ less than 1.5 was expected to be observed in real samples with a very small amount of impurities even for the case of a three-dimensional antiferromagnet~\cite{rosch1,rosch2}. The behavior of the exponent $n$ strongly depends on the concentrations of a small amount of impurities. The experimental data of CePd$_2$Si$_2$ was discussed on this point of view~\cite{rosch1,grosche}. However, this model does not seem to be consistent with the present case of CeIrSi$_3$ from a following fact that two samples having different residual resistivity show almost same exponent $n$ = 1.

   The $T$-linear dependence of the resistivity is also predicted by the critical valence fluctuation (CFV) mechanism of unconventional superconductivity in Ce compounds as CeCu$_2$Si$_2$ and CeCu$_2$Ge$_2$~\cite{miyake1,miyake2,watanabe,holmes1,holmes2}.  According to the theory, a sharp valence change is caused by the strong and local Coulomb repulsion $U_{\rm cf}$ between $f$ and conduction electrons and the superconducting state with the $d$-wave symmetry is induced by the process of exchanging the slave-boson fluctuations. The $T$-linear dependence of the resistivity was predicted to appear in a small parameter (pressure) region around the critical valence transition at $P$ = $P_{\rm v}$.  One may consider that the superconductivity in CeIrSi$_3$ is mediated by the CVF and the critical pressure $P_{\rm v}$ locates around $2.5\,-\,2.6$ GPa where the superconducting heat capacity anomaly ${\Delta}{C_{\rm ac}}/C_{\rm ac}(T_{\rm sc})$ and $T_{\rm sc}$ show maximum values. However, contrary to the theoretical expectation, the residual resistivity ${\rho}_0$ and the linear heat capacity coefficient $\gamma$ are not enhanced around the critical pressure.

\subsection{Anomalous temperature dependence of the upper critical field  $B_{c2}$}  
  Figure 7 shows the temperature dependence of the upper critical field $B_{\rm c2}$ at 2.65 GPa. The data are cited from our previous paper~\cite{okuda}. The upper critical field is highly anisotropic: $-{\rm d}B_{\rm c2}$/d$T$ = 14.6 T/K at $T_{\rm sc}$ = 1.58 K for $B$ $\parallel$ [001], and $-{\rm d}B_{\rm c2}$/d$T$ = 13.0 T/K at $T_{\rm sc}$ = 1.62 K and $B_{\rm c2}(0)$ = 9.5 T for $B$ $\parallel$ [110]. The upper critical field $B_{\rm c2}$ for $B$ $\parallel$ [110] indicates the tendency of the Pauli paramagnetic suppression, while the upper critical field for $B$ $\parallel$ [001] is not destroyed by spin polarization based on the Zeeman coulpling and possesses an upturn curvature below 1 K. The upper critical field $B_{\rm c2}$  $\parallel$ [001] is extremely large and it is roughly estimated as 30 T. 
         \begin{figure}[t]
\begin{center}
\includegraphics[width=15cm]{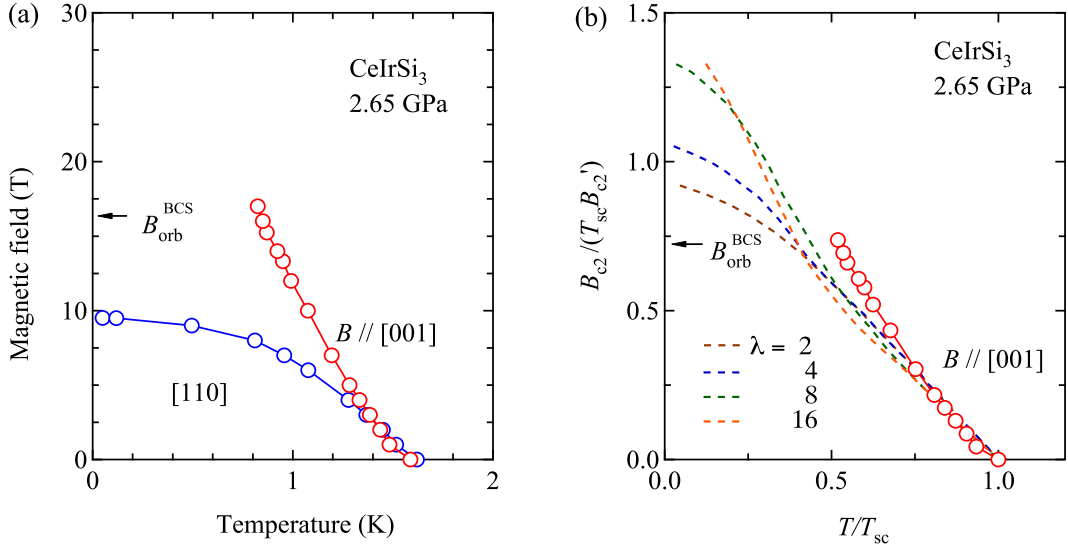}
 \end{center}
\caption{\label{fig:epsart} (a) Temperature dependence of the upper critical field $B_{\rm c2}(T)$ for $B$ $\parallel$ [110] and [001] at 2.65 GPa in CeIrSi$_3$~\cite{okuda}. (b) $B_{\rm c2}$($T$) curves for  $B$ $\parallel$ [001] normalized by the initial slope $B'_{\rm c2}$ and the superconducting transition temperature $T_{\rm sc}$. The arrow indicates the orbital limit $B_{\rm orb}^{\rm BCS}$ = 0.73 $B'_{\rm c2}$$T_{\rm sc}$ for  $B$ $\parallel$ [001]. The dashed curves are theoretical calculations based on the strong-coupling model using the coupling strength paramter $\lambda$ =2, 4, 8 and 16~\cite{bulaevskii}.}
\end{figure}     

  Superconductivity in the non-centrosymmetric crystal structure was theoretically discussed~\cite{frigeri}. The superconducting gaps for the spin-singlet and triplet channels are coupled by a finite value of the antisymmetric spin-orbit coupling $\alpha$ and thus the gap function is a mixture of both channels. Frigeri, Agterberg and Sigrist calculated the spin susceptibility (${\chi}_{\rm s}$) of both the singlet and triple components for the field directions parallel (${\chi}_{\rm s}^{\parallel }$) and perpendicular (${\chi}_{\rm s}^{\perp}$) to the $c$-axis for the case of the tetragonal non-centrosymmetric superconductor with the Rashba-type spin-orbit interaction where the potential gradient ${\nabla}V(\mathbf{r})$ appears along the $c$-axis~\cite{frigeri}. At $T$ = 0 K, the values of ${\chi}_{\rm s}^{\parallel}$ and ${\chi}_{\rm s}^{\perp}$ of the singlet component increases with the spin-orbit coupling strength $\alpha$ and  then ${\chi}_{\rm s}^{\parallel}$ and ${\chi}_{\rm s}^{\perp}$ approache the normal state spin susceptiblity ${\chi}_{\rm n}$ and ${\chi}_{\rm n}$/2, respectively. For the triplet component, ${\chi}_{\rm s}^{\parallel}$ and ${\chi}_{\rm s}^{\perp}$ are ${\chi}_{\rm n}$ and ${\chi}_{\rm n}$/2, respectively, at any $\alpha$. In both singlet and triplet components, the paramagnetic pair breaking effect is very anisotropic. There is almost no paramagnetic suppression for $B$ $\parallel$ [001] in the Rashba-type spin-orbit interaction, while the suppression exists for $B$ $\parallel$ [110]  (in plane) depending on the $\alpha$ value. The present large anistropy of $B_{\rm c2}$(0) in CeIrSi$_3$ is qualitatively consistent with the theoretical prediction~\cite{frigeri,okuda}.
 
 The upper critical field $B$ $\parallel$ [001] should be restricted by the orbital limitting field $B_{\rm orb}$ which is expressed as $B_{\rm orb}^{\rm BCS}$ = 0.73$B'_{\rm c2}$$T_{\rm sc}$ by the weak-coupling BCS theory~\cite{helfand}, even though the paramagnetic pair-breakng effect is almost absent for the direction. Here, $B'_{\rm c2}$ is the slope of the upper critical field at $T_{\rm sc}$, $-{\rm d}B_{\rm c2}$/d$T|_{T=T_{\rm sc}}$. The value of $B_{\rm orb}^{\rm BCS}$ is estimated as 15.1 T for $B$ $\parallel$ [001] which is obviously smaller than the experimental value as shown in Figure 7(a). Also, the temperature dependence of $B_{c2}(T)$ for $B$ $\|$ [001] shows an unusual temperature dependence, a positive curvature, which can not be explained by the weak-coupling BCS model. In order to explain these phenomena, the strong-coupling effect should be taken into account. Theoretically, the positive curvature of the orbital critical field $B_{\rm orb}$ is expected when the strong-coupling parameter $\lambda$ is large~\cite{bulaevskii}. The temperature dependence of $B_{c2}(T)$ in UBe$_{13}$ was analyzed from this point of view~\cite{thomas}. In CeIrSi$_3$, the Pauli paramagnetic effect for $B$ $\|$ [001] is strongly reduced and the orbital effect becomes dominant in the temperature dependence of $B_{\rm c2}(T)$ under low and moderate magnetic fields at low temperatures. The dashed curves shown in Figure 7(b) are results of the theoretical calculation by the strong-coupling theory without the paramagnetic pair-breaking effect for a clean limit superconductor~\cite{bulaevskii}. The model assumes a superconductor with the conventional electron-phonon type. Although it is supposed that the coupling of electrons in CeIrSi$_3$ is mediated by magnetic interactions rather than phonon, it is useful for understanding of the behavior of $B_{\rm c2}(T)$ in the case of strong-coupling superconductivity, regardless of the pairing mechanism. The positive curvature of the $B_{\rm c2}$-curve is roughly reproduced by the model but the data deviate from the theoretical curves roughly below $T/T_{\rm sc}$$<$ 0.8. A larger value of the coupling parameter seems to be needed. Also, it should be noted that the model assumes that the spherical Fermi surface and the electron-phonon coupling for the superconductivity as mentioned before. To reproduce the data more precisely, we must take into account more detailed Fermi surface topology and the paring mechanism.

  As we have discussed in this section, the present study suggests that the large magnitude and anisotropy of $B_{c2}$(0) in CeIrSi$_3$ is a result of combined effects of the strong-coupling superconductivity and the reduced paramagnetic effect of the non-centrosymmetric superconductor. The similar large magnitude and anisotropy of $B_{c2}$(0) was also reported in CeRhSi$_3$ and analyzed on the same point of view~\cite{kimura3}. In these analyses, the spin susceptiblity $\chi_{\rm s}$ is assumed to be isotropic and the orbital part (Van Vleck susceptibility) $\chi_{\rm orb}$ is neglected. The anisotropy of the magnetic susceptiblity ${\chi}_{[110]}/{\chi}_{[001]}$ at ambient pressure is about 2 at low temperatures.  As we discussed in the section {\it 3.4}, the anisotropy of the magnetic property does not seems to change significantly under high pressure and the anisotropy is not enough to explain that of $B_{c2}$(0)~\cite{okuda}. In the case of heavy-fermion system, the orbital part is usually large, compared the spin part\cite{tou}. For further quantitative analysis on $B_{c2}$(0), it is necessary to estimate the spin susceptibility from the NMR experiment.

  \subsection{Comparison with other Ce-based heavy-fermion superconductors}  
  We compare the present experimental results with those of other Ce-based heavy-fermion superconductors. There are two categories for the superconductors characterized by the shape of their superconducting region in the pressure-temperature phase diagram. For the details of this classification, the readers refer to the ref. 59~\cite{holmes2}. The first category contains pressure induced superconductors such as CeIn$_3$ or CePd$_2$Si$_2$ whose small superconducting phase is situated around the magnetic critical pressure $P_{\rm c}$~\cite{jaccard,mathur}. In the second category, a superconducting phase is found over a much broader  pressure range than in the first category, extending far from $P_{\rm c}$. The superconducting transition temperature $T_{\rm sc}$ shows a maximum value at the pressure higher than $P_{\rm c}$. CeIrSi$_3$ seems to belong to the second category. The present experimental results in CeIrSi$_3$ are discussed in comparison with those of CeRhIn$_5$, CeCoIn$_5$ and CeCu$_2$Si$_2$ which belong to the second category~\cite{petrovic,ikeda,hegger}.  In CeRhIn$_5$,  the superconductivity appears above the antiferromagnetic critical pressure $P_{c}^{*}$ = 1.95 GPa~\cite{hegger}. The cyclotron effective mass obtained from the de Haas-van Alphen experiment and the residual resistivity $\rho_{0}$ indicate a divergent tendency around the critical pressure $P_{\rm c}$ = 2.35 GPa where $T_{sc}$ shows a maximum value~\cite{muramatsu,shishido}. A marked change in the 4{\it f} electron nature from localized to itinerant states is realized at $P_{\rm c}$ under magnetic field. The value of ${\Delta}{C_{\rm ac}}/C_{\rm ac}(T_{\rm sc})$ shows a maximum value of 1.42 at $P_{\rm c}$~\cite{knebel}. In CeCoIn$_5$ where the antiferromagnetic critical pressure is located at the negative pressure side~\cite{pagliuso}, the large specific heat jump (${\Delta}{C}/{\gamma}{T_{\rm sc}}$ = 4.5) was observed at $T_{\rm sc}$. Under high pressure, ${\Delta}{C_{\rm ac}}/C_{\rm ac}(T_{\rm sc})$ decreases with increasing pressure from 4.5 at 0 GPa to 1.0 around 3 GPa~\cite{knebel}. Correspondingly, the values of  $\gamma$ and $\rho_{0}$ decrease considerably~\cite{sparn,sidorov}. High pressure experiment on CeCu$_2$Si$_2$ clarified that $T_{\rm sc}$ is enhanced around 4 GPa where the residual resistivity $\rho_0$ and the superconducting heat capacity jump ${\Delta}{C}/{\gamma}{T_{\rm sc}}$ also show maximum values~\cite{holmes1}. In these three superconductors, even if the origin for the enhancements of the physical quantities may differ in each compound, the jump of the heat capacity at $T_{\rm sc}$, ${\Delta}{C_{\rm ac}}/C_{\rm ac}(T_{\rm sc})$, correlates with the enhancements of the $\gamma$ and $\rho_{0}$ values. On the other hand, in CeIrSi$_3$, no divergent tendency is observed in $\gamma$ and $\rho_{0}$ at $P_{\rm c}$ = 2.25 GPa and around $2.5-2.7$ GPa, where ${\Delta}{C_{\rm ac}}/C_{\rm ac}(T_{\rm sc})$ shows a maximum value. Theoretically, the non-centrosymmetric superconductivity of CeIrSi$_3$ needs to be considered, especially on the basis of the present experimental result that CeIrSi$_3$ is a strong-coupling superconductor with a moderate value of $\gamma$ = 100 $\pm$ 20 mJ/K${^2}{\cdot}$mol. 

  Finally,  we discuss the present results from the view point of a theoretical interpretation of the large jump of the heat capacity at $T_{\rm sc}$ in CeCoIn$_5$ and UBe$_{13}$ by Kos, Martin and Varma~\cite{varma}. The authors claimed that the large value of ${\Delta}{C}/{\gamma}{T_{\rm sc}}$ in both compounds is not strong-coupling effect but is caused by the coupling between the superconducting ordering parameter $\psi$ and fluctuating magnetic moments of localized $f$ electrons~\cite{varma}. The values of ${\Delta}{C}/{\gamma}{T_{\rm sc}}$ in CeCoIn$_5$ and UBe$_{13}$ become considerably small if the enhanced value of $C/T$ under magnetic field above the upper critical field is used as the ${\gamma}$ value, which is not understood as the strong-coupling superconductor described by the Eliashberg theory~\cite{flouquet,petrovic,varma}. The coupling between the superconducting order paramater $\psi$ and fluctuation magnetic moments decreases the superconducting transition temperature $T_{\rm sc}$ and increases the value of ${\Delta}{C}/{\gamma}{T_{\rm sc}}$. Indeed, experimental data of CeCoIn$_5$ under high pressure are consistent with the theory.  
In the theory, a low energy scale $T_{Fl}$ is introduced in a similar way to  ``Two-fluid model'' for the Kondo lattice\cite{nakatsuji1}. $T_{Fl}$ is assumed to be much lower than $T_{\rm sc}$. The large value of ${\Delta}{C}/{\gamma}{T_{\rm sc}}$ arises from the coupling of the order parameter $\psi$ and the magnetic fluctuations when the latter can be treated classically near $T_{\rm sc}$ ($T_{Fl}$ $\ll$ $T_{\rm sc}$). A key point is whether the low energy scale ($T_{Fl}$) does exist or not. In CeCoIn$_5$, a characteristic energy scale of 1.7 K has been deduced from the specific heat measurement above the upper critical field and a systematic study on the La dilution Ce$_{1-x}$La$_x$CoIn$_5$ system\cite{petrovic, varma, nakatsuji1,nakatsuji2}. This value is close to a ``low Kondo temperature" $T_{\rm K}$ = 1.5 which is obtained by a theoretical expression $k_{\rm B}T_{\rm K} = (k_{\rm B}T^h_{\rm K})^3/{\Delta_1}{\Delta_2}$\cite{yamada}. Here, $T^h_{\rm K}$ is the high Kondo Temperature. ${\Delta_1}$ and ${\Delta_2}$ are widths of the CEF splitting. In order to check this point, the precise measurement of the heat capacity above the upper critical field $B_{\rm c2}$ is needed in CeIrSi$_3$,  which is not available for the moment. We have $T^h_{\rm K}\,{\sim}\,100$ K,  ${\Delta_1}$ = 149 K, and ${\Delta_2}$ = 462 K for CeIrSi$_3$\cite{okuda}. The low Kondo temperature is estimated as 12 K using the theoretical expression. The value is much higher than $T_{\rm sc}$. The strong-coupling effect is obvious not only from the the large value of ${\Delta}{C_{\rm ac}}/C_{\rm ac}(T_{\rm sc})$ but also from the temperature dependence of $B_{\rm c2}(T)$ for $B$ $\parallel$ [001] at 2.65 and the pressure dependence of $-dB_{\rm c2}/dT$ at $T_{\rm sc}$, as we discussed above.  It is noted that the pressure dependences of $T_{\rm sc}$ and ${\Delta}{C_{\rm ac}}/C_{\rm ac}(T_{\rm sc})$ in CeIrSi$_3$ shows an opposite tendency to CeCoIn$_5$ and the theoretical prediction. The situation in CeIrSi$_3$ seems to be different from the case of CeCoIn$_5$.

\section{Conclusions}
We have studied the pressure-induced superconductor CeIrSi$_3$ by heat capacity and electrical resistivity measurements under high pressure. The experimental results are summarized as follows.

  1) The critical pressure of the antiferromagnetic state is determined to be $P_{\rm c}$ = 2.25 GPa. Bulk superconductivity is mainly realized above $P_{\rm c}$. It seems that the antiferromagnetism and superconductivity essentially do not co-exist below $P_{\rm c}$. The electrical current $j$ dependence of the superconducting transition temperature $T_{\rm sc}$ and the disappearance of the superconductivity at higher electrical current $j$ indicate the spatially inhomogeneous superconductivity at 1.31 and 1.71 GPa. 
  
  2) The highest $T_{\rm sc}$ = 1.6 K and ${\Delta}{C_{\rm ac}}/C_{\rm ac}(T_{\rm sc})$ = 5.7 $\pm$ 0.1 values are obtained at pressures higher than $P_{\rm c}$, namely, around $2.5-2.7$ GPa. The value of ${\Delta}{C_{\rm ac}}/C_{\rm ac}(T_{\rm sc})$ is the largest value among all superconductors. The present observation indicate that CeIrSi$_3$ is a strong-coupling superconductor. 

  3) The $\gamma$ value of 100 $\pm$ 20 mJ/K$^2$$\cdot$mol at $T_{\rm sc}$ is approximately unchanged as a function of  pressure. There is no anomalous enhancement in the pressure dependences of the linear heat capacity coefficient $\gamma$ and residual resistivity ${\rho_0}$. This suggests that the magnetic critical pressure $P_{\rm c}$ is not second order quantum critical point. The superconductivity may be different from those appeared around the magnetic instability.

  4) Above $P_{\rm c}$, the temperature dependence of the resistivity shows the anomalous $T$-linear dependence. Meanwhile, the heat capacity $C_{\rm ac}/T$ show a monotonic and weak temperature dependence in the normal state above $T_{\rm sc}$. These behaviors can not be explained simply by the spin fluctuation theory for the three-dimensional antiferromagnet.
       
  5)  The large magnitude and anisotropy of the upper critical field $B_{c2}$ is a result of combined effects of the strong-coupling superconductivity and the reduced paramagnetic effect of the non-centrosymmetric superconductor. The strong-coupling effect on the orbital critical field reflects the downward curvature in the temperature dependence of $B_{\rm c2}(T)$ for $B$ $\parallel$ [001] at 2.65 GPa.

  6) The jump of the heat capacity at $T_{\rm sc}$, ${\Delta}{C_{\rm ac}}/C_{\rm ac}(T_{\rm sc})$ does not correlate with the enhancements of the $\gamma$ and $\rho_{0}$ values. This is contrary to the cases of other Ce-based heavy-fermiom superconductors CeCu$_2$Si$_2$, CeRhIn$_5$ and CeCoIn$_5$.
       
 \section*{Acknowledgements}
This work was financially supported by the Grant-in-Aid for Creative Scientific Research (15GS0213), Scientific Research of Priority Area and Scientific Research (A and C) from the Ministry of Education, Culture, Sports, Science and Technology (MEXT).

\smallskip


\end{document}